\documentclass[12pt,a4paper]{article}
\usepackage{graphicx}
\begin{document}
 
 \thispagestyle{empty}
 
 \title{Quasiperiodic propagation in time of some classical/quantum systems:
 Nielsen's conserved quantity and Floquet properties.}
 \author{P Kramer, Theoretische Physik U. Tuebingen, Germany,\\ 
 T Kramer, Institute for Theoretical Physics, U. Regensburg, Germany,\\ V  I Man'ko, Lebedev Physical Institute, Moscow,
 Russia.}
 \maketitle

\section*{Abstract.}
We consider classical and quantum propagators for two different time intervals. If these 
propagators follow one another in a Fibonacci sequence we get a discrete quasiperiodic system.
A theorem due to Nielsen provides a novel conserved quantity for this system.
The Nielsen quantity controls the transition between  commutative and non-commutative 
propagation in time. The quasiperiodically kicked oscillator moreover 
is dominated by  quasiperiodic analogues of the Floquet theorem.

\section{Introduction.}

For the general description  of time-dependent systems in quantum mechanics 
we refer to \cite{DO}, \cite{MM}, \cite{KA1}, \cite{KA2}, \cite{BY}.
We consider the time evolution for a  particular class of 
time-dependent classical/quantum systems which are periodic
or quasiperiodic in time. 

We restrict our attention to time evolutions,  built 
piecewise from strings with classical symplectic propagators $g(t)$
in $2D$ phase space.
These describe oscillators and dilatations, or positive and negative 
$\delta$-kicks.  Quantum propagators are
constructed as unitary representations $S_{g(t)}$ of the classical
linear canonical transformations  for $g(t)$ from 
the symplectic group $Sp(2,R)$.

We present Nielsen's theorem 
on automorphims of the free group $F_2$ with
two generators and  from it derive a conserved 
quantity in terms of  the group commutator. We 
analyze Floquet properties of classical and 
quantum propagators. 
We show  the presence/absence  of
Floquet quantum numbers for periodic system under the control of 
system parameters.
For the periodically kicked classical oscillator we demonstrate the 
occurrence of Floquet bands and Floquet gaps.

For periodic systems built from two basic intervals, we construct Nielsen's conserved 
quantity as a function of the system parameters.    
We extend the analysis to quasiperiodic strings.
By building the strings  through a Fibonacci automorphism
we are able to survey their quasi-Floquet properties.

\section{Floquet theorem for linear differential equations.}

{\bf 1 Theorem Floquet}: Consider the real first-order periodic linear system of deq.
\begin{equation}
\label{f1}
\frac{d}{dt} Y(t)= A(t) Y(t),\: A(t)\; {\rm real},\: A(t+T)=A(t), \; T>0.
\end{equation}
A fundamental system of two solutions can be written in the form
\begin{equation}
\label{f2}
Y(t) = \exp(\pm iQt) Y_0(t),\; Y_0(t+T)=Y_0(t),\; 0\leq Q <\frac{2\pi}{T}.
\end{equation}
We denote the real number $Q$ as the Floquet index and $\exp(iQT)$
as the Floquet factor. It will become clear in what follows that the 
Floquet theorem requires complex solutions of the deq eq.\ref{f1}
for real $A(t)$.

{\em Proof}: The discrete time translation   $T: A(t)\rightarrow A(t+T)$ 
is a symmetry operation of $A(t)$. It generates the abelian group $C_{\infty}$
with elements $\pm nT, n=0,1,...$. The unitary representations of 
this group have the form 
\begin{equation}
\label{f3}
T \rightarrow \exp(i\Omega T),\; D^{\Omega}(nT)= \exp(in\Omega T), 
0\leq \Omega <\frac{2\pi}{T}.
\end{equation} 
The orthogonality and completeness relation of $C_{\infty}$ are given 
in Appendix A.

Consider a  solution of eq. \ref{f2} which can be arranged so that it transforms 
according to an irreducible representations of $C_{\infty}$,
\begin{equation}
\label{f4}
Y^{\Omega}(t): Y^{ \Omega}(t+T)= Y^{\Omega}(t)\exp( i\Omega T).
\end{equation}
>From this solution define the function
\begin{equation}
\label{f5}
Y_0^{ \Omega} (t):= \exp( -i\Omega t)Y^{ \Omega}(t)
\end{equation}
Then it easily follows that $Y_0^{ \Omega}(t)$ is periodic,
\begin{equation}
\label{f6}
Y_0^{ \Omega} (t+T)=Y_0^{ \Omega} (t)
\end{equation}
Since $A(t)$ is real, 
the complex conjugate solution 
$\overline{Y^{\Omega}}:=Y^{- \Omega}$ is  
a second linearly independent  solution with the complex conjugate Floquet factor. 
With $\Omega=Q$ we obtain eq. \ref{f2}.

Remark (i):The Floquet theorem is mathematically equivalent to the Bloch 
theorem for systems with discrete position symmetry, which has the same
abstract symmetry group $C_{\infty}$. 

Remark (ii): We remark that the periodic system eq. \ref{f1} for certain ranges 
of its parameters cannot have solutions with the Floquet property. 
We shall give examples  of this 
situation in later sections.

\section{The free group $F_2$ and its automorphisms.}

We refer to \cite{MA} for what follows. We also quote \cite{ML} 
as a useful reference to basic concepts for free monoids.

{\bf 2 Def}. The infinite {\em free group} $F_2$ consists of all
words generated by concatenation from  two non-commuting invertible  generators 
$\{y_1, y_2\}$. Its 
elements can be viewed as a tree formed by words of increasing length.

{\bf 3 Def}: An automorphism of $F_2$ is an invertible homomorphic map 
\begin{equation}
\label{ff0}
\phi: F_2 \rightarrow F_2,
\; \{y_1,y_2\} \rightarrow \{ \phi_1(y_1,y_2), \phi_2(y_1, y_2)\},
\; \{\phi_1, \phi_2\} \in F_2,
\end{equation}
and so preserves group multiplication and unit element.
The set of all automorphisms of $F_2$ forms the infinite group
$\{\phi\} = Aut(F_2)$. This group, as was shown by Nielsen \cite{MA},
is again finitely generated, with explicit 
generators given in \cite{MA}. Similar results hold true for the 
free group $F_n$ and its group of automorphims $Aut(F_n)$. 
Homomorphic properties of finitely generated groups have the virtue that they 
can be verified by checking them only 
for the generators. 

{\em Example 1}: The Fibonacci automorphism is defined by
\begin{eqnarray}
\label{ff1}
&&\phi_{fibo}: \{y_1, y_2\} \rightarrow 
\{\phi_{fibo,1}(y_1, y_2),\phi_{fibo,2}(y_1, y_2)\}: = \{y_2, y_1y_2\},
\\ \nonumber
&& \phi_{fibo}^{-1}:   
\{y_1, y_2\} \rightarrow \{y_2y_1^{-1}, y_1\},
\end{eqnarray}
We shall use in particular iterated automorphims as $(\phi_{fibo})^n$.
Then the words $(\phi_{fibo})^n(y_1)$, with their length $|w|$,
defined \cite{ML} as the number 
of letters in $w$ after reducing $w$ by use of all relations
of the type $yy^{-1}=e$,  become
\begin{equation}
\label{ff1a}
\begin{array}{lllllll} 
(\phi_{fibo})^n(y_1):& y_1& y_2& y_1y_2& y_2y_1y_2& y_1y_2^2y_1y_2&...\\
|(\phi_{fibo})^n(y_1)|:          & 1  &1   &2      &3         &5   &...
\end{array}
\end{equation}
The word length $|(\phi_{fibo})^j(y_1)|$ is the  Fibonacci number $n_j$
\begin{equation}
\label{ff1b}
n_j: n_{j+2}=n_j +n_{j+1}, n_1=n_2=1. 
\end{equation}

Notice the recursive structure of the images under $(\phi_{fibo})^n$:
Define $(\phi_{fibo})^{(n-1)}(y_1)=:w_1$,
$(\phi_{fibo})^{(n-1)}(y_2)=:w_2.$ Then
it follows from eq. \ref{ff0} that 
$(\phi_{fibo})^n(y_1)= w_2$, $(\phi_{fibo})^n(y_2)=(w_1w_2)$, and so 
we must in step $n$ simply concatenate the two words $(w_1,w_2)$ 
obtained in step $(n-1)$.

{\bf 4 Nielsen theorem on $Aut(F_2)$}:
Consider the commutator ${\cal K}=y_1y_2y_1^{-1}y_2^{-1}$ in $F_2$ and a fixed 
automorphism $\phi(F_2) \in Aut(F_2)$. Then the image of the commutator
under $\phi$ obeys, \cite{MA} theorem 3.9 p.  165, 
\begin{equation}
\label{ff2}
\phi({\cal K})=w {\cal K}^{\pm 1}w^{-1},\; w \in F_2,
\end{equation}
where the sign $\pm 1$ and the element $w$ depend on the chosen 
automorphism $\phi$.

{\em Proof}: It suffices to prove the result for the finite set of
generators of $Aut(F_2)$ given by Nielsen \cite{MA}.

So the commutator ${\cal K}$ is conserved, up to inversion and conjugation,
under any automorphism of $Aut(F_2)$.
No counterpart of the Nielsen theorem 3 is known for $Aut(F_n), n>3$.

{\em Example 2}: Evaluation of the image of the commutator 
under the particular Fibonacci automorphism $\phi_{fibo}$ yields
\begin{equation}
\label{ff3}
{\cal K}:=y_1y_2y_1^{-1}y_2^{-1},\:  
\phi_{fibo}({\cal K})= y_2(y_1y_2)y_2^{-1}(y_1y_2)^{-1}
=y_2y_1y_2^{-1}y_1^{-1}={\cal K}^{-1}.
\end{equation}

The power of the Nielsen theorem  appears through the notion of
automorphisms induced on some group $G$.

{\bf 5 Def}: Induced automorphism: Consider a homomorphism $hom: F_2 \rightarrow G$ defined in terms
of the generators by $hom: \{y_1, y_2\} \rightarrow \{g_1, g_2\} \in G$.
The automorphism of $G$ induced by the automorphism $\phi \in Aut(F_2)$
eq. \ref{ff1} is the map $\phi_G: \{g_1, g_2\} \rightarrow 
\{\phi_1(g_1, g_2), \phi_2(g_1, g_2)\}$.
Clearly induced automorphisms on $G$ are homomorphic to the corresponding 
automorphisms from $Aut(F_2)$. 

For an induced automorphism, the commutator induced in the group $G$
has additional significance since it measures the non-commutativity of
the group. In case ${\cal K}=I$ we have commutativity on the induced level
which allows to rearrange the order of successive propagators.

{\em Example 3}: We shall consider in particular automorphisms induced
on the symplectic group $Sp(2,R)$, the group of linear canonical transformations
of a 1D classical hamiltonian system. For $g \in Sp(2,R)$, we have 
the algebraic equivalence relation $g^{-1} \sim g$ and 
the trace relation ${\rm tr}(g^{-1})={\rm tr}(g)$. Note that the character
is a class function. If these properties are combined with the Nielsen theorem,
one finds 

{\bf 6 Prop}: For any automorphism of the free group $F_2$ induced on $Sp(2,R)$,
the half-trace of the commutator,
\begin{equation}
\label{ff4}
\frac{1}{2} {\rm tr}({\cal K}) 
= \frac{1}{2} {\rm tr} (g_1g_2g_1^{-1}g_2^{-1})
\end{equation}
is a  quantity conserved under  iterated automorphisms.

The condition $\frac{1}{2}\chi ({\cal K})=1$ is a necessary but not a sufficient
condition for ${\cal K}=e$. So commutativity requires extra checking. 

If by iterated automorphisms  we can generate a string  of arbitrary length,
the conserved quantity eq. \ref{ff4} can be assigned 
to the infinite string.

If we choose the Fibonacci automorphism eq. \ref{ff1} and associate two matrix propagators
to its two intervals, we generate a quasiperiodic system. We term such a system discrete 
quasiperiodic. Such  system  form a special class among general 
quasiperiodic systems \cite{PK4}.

In  later sections, we shall associate to the classical system 
a quantum system. To a symplectic matrix $g$ 
we shall associate a representation
by a unitary operator $U(g)$ in Hilbert space. 

Since this association is
a homomorphic map from symplectic matrices 
into propagators which represent linear canonical transformations, 
the Nielsen theorem then extends
to a theorem in Hilbert space.

\section{Geometry and dynamics of systems of traces.}

The traces of the symplectic matrices $g \in Sp(2,R)$ obviously 
play an important
part in the classical and  quantum analysis. For a subclass of systems, the computation of
these traces does not even require the computation of the underlying matrices
\cite{PK2}. The Fibonacci system Example 1 belongs to this class. The 
trace invariant eq. \ref{ff4} geometrically becomes a cubic surface in 3-space
whose shape depends on the value of the invariant.
The traces under repeated automorphisms and with 
increasing string length determine recursively a discrete dynamical
system on this conserved surface. We sketch this approach:
For the Fibonacci system eq. \ref{ff1} we define 
$y_3:=y_1y_2$ and $g_3:=g_1g_2$.
With the induced matrix automorphism written as
\begin{equation}
\label{ff3a}
(X^n, Y^n,Z^n):=(\phi^n(g_1),\phi^n(g_2),\phi^n(g_3))
\end{equation}
and the matrix decomposition
\begin{equation}
\label{ff3b}
g \in Sp(2,R):\; g=S(g)+A(g),\; S=\frac{1}{2}{\rm trace}(g)e,\; A=g-S, {\rm trace}(A)=0
\end{equation}
one finds from eqs. (57-63) of \cite{PK2} the recursion relations
and the conserved quantity
\begin{eqnarray}
\label{ff3c}
&&\left[
\begin{array}{l}
S(X^{(n+1)})\\
S(Y^{(n+1)})\\
S(Z^{(n+1)})
\end{array}
\right]
=\left[
\begin{array}{l}
S(Y^n)\\
S(Z^n)\\
2S(Y^n)S(Z^n)-S(X^{n})
\end{array}
\right]
\\ \nonumber
&&S({\cal K})=-4S(g_1)S(g_2)S(g_3)
+2\left[(S(g_1)S(g_1)+S(g_2)S(g_2)+S(g_3)S(g_3)\right]-e
\end{eqnarray}
All the quantities in eq. \ref{ff3c} by eq. \ref{ff3b} can be expressed in terms
of real traces.
The recursion relation in eq. \ref{ff3c} determines a discrete dynamical 
system with  initial data given by the traces of $(g_1,g_2,g_3)$.
By use of eq. \ref{ff3b}, the conserved quantity 
$\frac{1}{2}\chi({\cal K})$ can be expressed in terms 
of the three initial traces. We refer to \cite{KO} for a view of
the cubic surface which represents $\frac{1}{2}\chi({\cal K})$.

\section{Quadratic hamiltonians and propagators.}

The quantum propagators for quadratic hamiltonians are given in \cite{TK1},
\cite{TK2}. They are unitary integral 
operators $S_{g(t_2-t_1)}$ where $g(t)$ is a linear canonical transformation
$g \in Sp(2m,R)$, which can be determined by exponentiation
from the action of the hamiltonian on the canonical position and momentum 
operators. So for quadratic hamiltonians we have 
a very clear correspondence between classical and quantum propagation:
The classical time evolution is given by the action of $g(t)$ on the
initial canonical coordinates at time zero. The quantum propagator is given by the 
action of $S_{g(t)}$ on the initial state at time zero.

\section{Hamiltonians and propagators for $n=1$.}

In \cite{TK1},\cite{TK2} one finds for quadratic hamiltonians the matrix analysis
of propagators in various dimensions $m$. The classification of
types of hamiltonians for $n=1,2$ is discussed in detail in \cite{WM}.
We briefly summarize the well-known results for $m=1$.

For $m=1$, that is for 1-dimensional position space, we have $g(t) \in Sp(2,R)$.
This group is particularly transparent. Its classes admit a subdivision into three 
types, depending on the value of their half-trace 
$ \frac{1}{2}\chi(g):=\frac{1}{2} tr(g)$. Since ${\rm det}(g)=1$, the eigenvalues
of $g$ are completely determined by the trace. Moreover the trace is independent 
of symplectic similarity transformations, $\chi(qgq^{-1})=\chi(g)$, 
and so allows to determine the class 
of the symplectic matrix $g$.
For 
$I: \frac{1}{2}|\chi(g)| <1$, $g$ is symplectically equivalent to the dynamics 
generated by an oscillator hamiltonian,
\begin{equation}
\label{f7}
I: H= \frac{1}{2m} P^2+\frac{m\omega}{2}X^2,
\end{equation}
for $II: \frac{1}{2}|\chi(g)| >1$ they are equivalent to an inverted oscillator or a
dilatation
\begin{equation}
\label{f8}
II: H= \kappa\frac{1}{2}(PX+XP)=\kappa(XP+\frac{h}{2i}),
\end{equation}
and for
$III:\frac{1}{2} |\chi(g)| =1$ they are, apart from the extra case $g=e,\; \frac{1}{2}\chi(e)=1$, 
equivalent to a free motion,
\begin{equation}
\label{f9}
III: H= \frac{1}{2m} P^2,
\end{equation}
All three hamiltonians are hermitian and time-independent. Hence 
their propagators $S_g(t)$ are unitary operators. For type I and III
the propagators can be written as integral operators according to 
Moshinsky and Quesne \cite{MQ} in the Schroedinger representation or 
in the Bargmann space of analytic functions, compare \cite{TK1}. 

{\em Type I}: Harmonic oscillator.

Rewrite the harmonic oscillator hamiltonian eq. \ref{f7} as
\begin{equation}
\label{f10}
H=\frac{1}{2}\omega \left[\frac{1}{m\omega}P^2+m\omega X^2\right].
\end{equation}
The corresponding linear canonical transformation is obtained as
\begin{eqnarray}
\label{f11}
&&g_{\omega}(t) =\left[
\begin{array}{ll}
\frac{1}{\sqrt{m\omega}}&0\\
0&\sqrt{m\omega}
\end{array}
\right]
\left[
\begin{array}{ll}
\cos(\omega t)&\sin(\omega t)\\
-\sin(\omega t)&\cos(\omega t)
\end{array}
\right]
\left[
\begin{array}{ll}
\sqrt{m\omega}&0\\
0&\frac{1}{\sqrt{m\omega}}
\end{array}
\right],
\\ \nonumber
&&
=\left[
\begin{array}{ll}
\cos(\omega t)&\frac{1}{m\omega}\sin(\omega t)\\
-(m\omega)\sin(\omega t)&\cos(\omega t)
\end{array}
\right]
\end{eqnarray}
Eq. \ref{f11} admits the complex diagonalization
\begin{eqnarray}
\label{f12}
&&\left[
\begin{array}{ll}
\cos(\omega t)&\sin(\omega t)\\
-\sin(\omega t)&\cos(\omega t)
\end{array}
\right]
\\ \nonumber
&&=\sqrt{\frac{1}{2}}
\left[
\begin{array}{ll}
1&i\\
1&-i
\end{array}
\right]
\left[
\begin{array}{ll}
\exp(-i\omega t)&0\\
0&\exp(i\omega t)
\end{array}
\right]
\sqrt{\frac{1}{2}}
\left[
\begin{array}{ll}
1&1\\
-i&i
\end{array}
\right]
\end{eqnarray}
In the complex domain we can therefore pass to a pair of 
classical variables which
propagate with $\exp(\mp i\omega t)$. 

The corresponding quantum propagator in the Bargmann space of 
analytic functions is \cite{TK1}
\begin{equation}
\label{f13}
\langle z'|S_{g_c(t)}|z\rangle
= \exp(-i\omega t/2)\exp(\exp(i\omega t) z' \overline{z}).
\end{equation}
The propagator in the Schroedinger representation is given in \cite{TK1}
pp. 34-6.

{\em Type II}: The hyperbolic and dilatation hamiltonian. 

For type II the symplectic matrix may be taken as
\begin{equation}
\label{f13a}
g(t)=
\left[
\begin{array}{ll}
\exp(\kappa t)&0\\
0&\exp(-\kappa t)
\end{array}
\right]
\end{equation}
Note the correspondence to eq. \ref{f12} for $\kappa \rightarrow -i\omega$.

There is for type II no
obvious integral operator representation. 
In this case we proceed directly from the
hamiltonian eq. \ref{f8} and find
\begin{equation}
\label{f14}
\exp(-\kappa t\frac{i}{h}H)\psi(x)=\exp(-\kappa \frac{t}{2})
(\exp(-\kappa tx\frac{d}{dx})\psi)(x)
=\exp(-\kappa \frac{t}{2})\psi(\exp(-\kappa t)x).
\end{equation}
This propagator is a time-dependent unitary dilatation.
Applied to the ground state of an oscillator with frequency $\omega_0$,
it transforms it into an oscillator ground state with frequency
$\omega(t)=\omega_0\exp(-2\kappa t)$. As a function of time, the uncertainty 
in position would increase while the uncertainty of momentum would 
decrease. 

Another form for class $II$ is given by the hyperbolic hamiltonian
\begin{equation}
\label{f14a}
H=\frac{1}{2}\left[\frac{P^2}{m}-m\kappa^2X^2\right].
\end{equation}
with the symplectic propagator 
\begin{equation}
\label{f14b}
g_{\kappa}(t)=
\left[ 
\begin{array}{ll}
\cosh (\kappa t)& (m\kappa)^{-1}\sinh(\kappa t)\\
(m\kappa)\sinh(\kappa t)& \cosh(\kappa t)
\end{array}
\right]
\end{equation} 
By choosing at time $t=0$ the initial matrix 
\begin{equation}
\label{B3}
h(0)= 
\left[ 
\begin{array}{ll}
1&0\\
0&m\kappa
\end{array}
\right], det(h(0))=m\kappa,
\end{equation}
we determine two canonical column solutions 
given by
\begin{equation}
\label{B4}
\left[ 
\begin{array}{ll}
X_I(t)&X_{II}(t)\\
P_I(t)&P_{II}(t)
\end{array}
\right]
=
g_{\kappa}(t) h(0)
=
\left[ 
\begin{array}{ll}
\cosh (\kappa t)&\sinh(\kappa t)\\
(m\kappa)\sinh(\kappa t)& (m\kappa) \cosh(\kappa t)
\end{array}
\right]
\end{equation}
Inserting the fundamental two solutions into the hamiltonian eq. \ref{f14a} 
we find the two different energies
\begin{equation}
\label{B5}
E_I(0)=-\frac{m\kappa^2}{2},\; E_{II}(0)=\frac{m\kappa^2}{2}.
\end{equation}
The types $I, II$ of propagation in phase space are shown in Fig. 1.

\begin{center}
\includegraphics{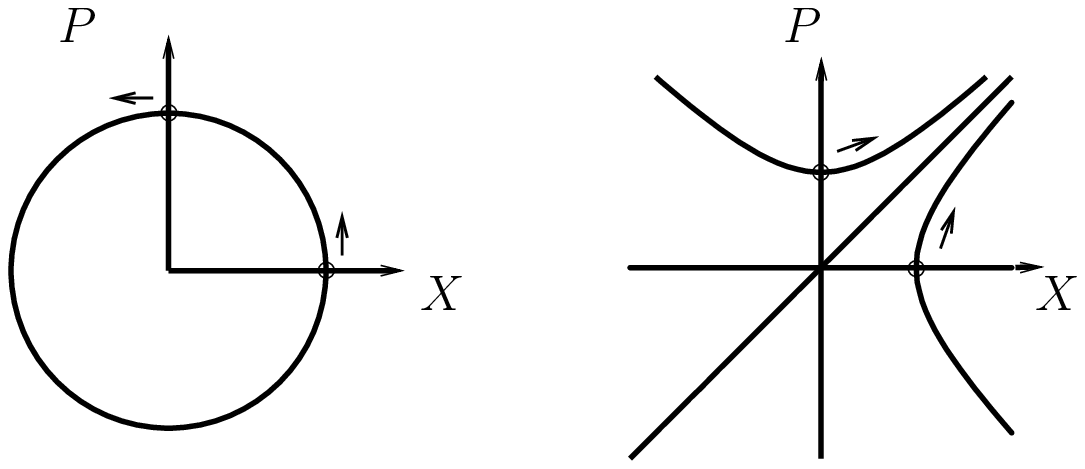}
\end{center}

Fig. 1. {\bf Phase space representation}. The  harmonic and the hyperbolic column 
solutions  I: eq. \ref{f11} and II: eq. \ref{f14b}  in phase space.
\vspace{0.2cm}

In appendix B we give the relation between the propagator and the 
transfer matrix.

{\em Type III}: Free particle.
>From the hamiltonian
\begin{equation}
\label{f15}
H=\frac{1}{2m}P^2
\end{equation}
one finds \cite{TK1} the symplectic matrix
\begin{equation}
\label{f16}
g(t)=
\left[
\begin{array}{ll}
1&-t\\
0&1
\end{array}
\right]
\end{equation}
and the propagator 
\begin{equation}
\label{f16a}
\langle x'|S_{g(t)}|x\rangle =
\exp(-i\pi/4)\sqrt{\frac{m}{2\pi ht}}\exp(i\frac{(x'-x)^2}{2t}).
\end{equation}

{\bf 7 Prop}: The types I, II, III together with the case $g=e$ exhaust, up to conjugation within
$Sp(2, R)$, the symplectic propagators with quadratic hamiltonians.

\section{Quadratic piecewise time-independent   hamiltonians.}

We consider now a system whose hamiltonian  for 
$0 \leq t \leq T= T_1+T_2$ is time-independent on the two intervals.  Assume that 
the propagation for the full interval $T$ is repeated periodically.  Over the period we have
then a propagation with the half-trace given by eq. \ref{l2}.

For $\frac{1}{2}|\chi|<1$ we can write $\frac{1}{2}\chi=\cos(\Omega T)$ and determine  
$\Omega, 0\leq \Omega < 2\pi/T$. This value $\Omega$ is the Floquet
index of a classical symplectic periodic system. The two Floquet factors
$\exp(\pm i\Omega T)$ appear as  the eigenvalues of the symplectic matrix
$g_1(T_1)g_2(T_2)$  
and determine  the classical propagators $U(T)$. The Floquet index
is completely determined by the parameters $(\omega_1,T_1,\omega_2,T_2)$
of the system. 
At multiples $nT, n=1,2,\ldots$ the classical propagator becomes
$(U(T))^n$.

In the quantum propagation,
$\Omega$  represents the overall frequency of an effective oscillator 
hamiltonian. The propagator over the period $T$ from eq. \ref{f13} becomes
\begin{equation}
\label{f19}
\langle z'|S_{g_c(T)}|z\rangle
= \exp(-i\Omega T/2)\exp(\exp(i\Omega T)z'\overline{z}), 
\end{equation}
We emphasize that this result holds true only for $T=T_1+T_2$ whereas
at intermediate time the propagator is governed by $g_1(t)$ or $g_2(t)$
respectively. The classical Floquet parameter enters the quantum propagator 
in an essential way. The eigenstates $|N\rangle$ of the overall harmonic oscillator
of frequency $\Omega$ from eq. \ref{f19} are the monomials of degree $N$.
These eigenstates under the discrete time translations transform according to 
\begin{equation}
\label{f19a}
T: |N\rangle \rightarrow |N\rangle \exp(iN\Omega T) 
\end{equation}
This again generates  an irreducible representation of the time translation group.
By the unique decomposition $N\Omega = m2\pi/T + \Omega', 0\leq \Omega' <2\pi/T,
m=0,1,2...$, this representation 
becomes $D^{\Omega}(NT)= D^{\Omega'}(T)$.

It follows that not the  quantum propagator itself, but only the eigenstates
in its decomposition transform irreducibly under the period $T$.
 
For $\frac{1}{2}|\chi|>1$ the situation changes. By writing 
$\frac{1}{2}|\chi|=\cosh (\kappa T)$
we determine an effective classical dilatation parameter $\kappa$. The eigenvalues
$\exp (\pm \kappa T)$ represent this dilatation over one period.
The corresponding classical propagation does not represent a Floquet 
propagation since the eigenvalues of the symplectic matrix do not 
have absolute value $1$. They form a non-unitary real representation 
of the time translation group.
The $n$th powers of these two factors now approach $0, \infty$ respectively.

The unitary quantum propagator over a period $T$  from eq. \ref{f14} becomes 
\begin{equation}
\label{f20}
\exp(-\kappa T\frac{i}{h}H)\psi(x)=\exp(-\kappa \frac{T}{2})
(\exp(-\kappa Tx\frac{d}{dx})\psi)(x)
=\exp(-\kappa \frac{T}{2})\psi(\exp(-\kappa T)x).
\end{equation}
Again, this propagator is valid only at time $T=T_1+T_2$. For intermediate time
the propagator is governed by one of the oscillator matrices $g_1(t)$ or
$g_2(t)$ respectively.
The quantum propagator over the periods $T$ stays unitary, but 
for $n \rightarrow \infty $ when acting on a state transforms it 
up to normalization  into a $\delta$-distribution
of sharp  effective momentum/position. 

Remark (i): The results on time-periodic systems from piecewise quadratic hamiltonians 
resemble the analysis of position-periodic Kronig-Penney systems \cite{KK}
with piecewise constant potentials. 
In the latter systems, the potentials play the role of the frequencies, and 
the Bloch $k$-label plays the role of the Floquet
index. Real values of the Bloch parameter yield a band index, while imaginary
values indicate   exponentially increasing or decreasing solutions 
of the Schroedinger equation which are excluded as band gaps.

{\em Example 5}: For the system of section 8 in case $|\frac{1}{2}\chi| <1$ we obtain 
an overall Floquet state with $\Omega$ determined by 
$\frac{1}{2}\chi= \cos (\Omega T)$.  Each Floquet state is the analogue 
to a Bloch state of a band model in $x$-space. 

\section{Oscillators and $\delta$-kicks.}

The transfer and propagator matrix methods apply to sequences of time intervals.

{\em Example 6}: The classical square well for the time interval $t$ 
has the type $I$ oscillator propagator
\begin{equation}
\label{zz1} 
g_{\omega}(t)=
\left[
\begin{array}{ll}
\cos(\omega t)&(m\omega)^{-1}\sin(\omega t)\\
-(m\omega)\sin(ka)&\cos(\omega t)
\end{array}
\right]
\end{equation}
The initial values for the two fundamental column solutions $I, II$ we 
take as 
\begin{equation}
\label{zz1a}
h(0)=
\left[
\begin{array}{ll}
1&0\\
0&m\omega
\end{array}
\right]
\end{equation}
Then the energies computed by insertion into the harmonic oscillator
hamiltonian become
\begin{equation}
\label{zz1b}
E_I=E_{II}= \frac{m \omega^2}{2}.
\end{equation}

{\em Example 7}: The square tunnel for the time  interval $t$ 
has the repulsive oscillator type $II$ propagator eq. \ref{f14b}.
The energy of the two fundamental solutions are
$E_I=-\frac{m\kappa^2}{2},  E_{II}=\frac{m\kappa^2}{2}$.

Following what is done in space in \cite{PK1}, \cite{PK2}  
we introduce 
w.r.t. time, negative and positive $\delta$-kicks. They are obtained as
limits  of square well and  square tunnel intervals of finite 
length. In the limits $: m\omega^2 T \rightarrow u',\; \omega T \rightarrow 0$
and $ m \kappa^2  T \rightarrow u',\; \kappa T \rightarrow 0$ 
with $u$ the strength parameter we
get from eqs. \ref{zz1}, \ref{zz1a} respectively the
symplectic transfer matrices for negative/positive $\delta$-kicks
in the form
\begin{equation}
\label{zz3} 
g_{\pm u'}=
\left[
\begin{array}{ll}
1 &0\\
\mp u'&1
\end{array}
\right].
\end{equation}

\section{The periodically kicked oscillator.}

We employ here the transfer matrix, see eq. \ref{B1}, with $u'=mu$,
\begin{equation}
 M(t)= 
\label{l1}
\lambda_i= 
\left[ 
\begin{array}{ll}
\cos(\omega T_i)& \omega^{-1}\sin(\omega T_i)\\
-\omega\sin(\omega T_i) & \cos(\omega T_i) 
\end{array}
\right]\:
\left[ 
\begin{array}{ll}
1& 0\\
u& 1 
\end{array}
\right],\;  i=1,2.
\end{equation}
with positive $\delta$-kicks. The half-trace becomes
\begin{equation}
\label{l2}
\chi= \frac{1}{2}{\rm Tr} (M(T))=\cos(\omega T)+\frac{u}{2\omega}\sin(\omega T) 
\end{equation}
This expression has the period $\frac{2\pi}{\omega}$. We find
\begin{equation} 
\label{l3}
\chi=0: \cot(\omega T_0)= -\frac{u}{2\omega} 
\end{equation}
The solutions $T_0$ of this equation exist in each interval $2m\pi  \leq \omega T \leq 2(m+1)\pi,\: m=0,\pm 1,\pm 2,..$. 
and mark the centers of Floquet bands as functions of $\kappa$.
The equations 
\begin{eqnarray}
 \label{l4}
&& \frac{1}{2}\chi=+1: \tan(\frac{\omega T_+}{2})=\frac{u}{2\omega},
\\ \nonumber 
&& \frac{1}{2}\chi=-1: \cot(\frac{\omega T_-}{2})=-\frac{u}{2\omega},
\end{eqnarray}
determine for any such band its edges $T_{\pm}$ as functions of $\kappa$.
Note that these equations have in $\omega T_{\pm}$ the period $4\pi$.
In between two Floquet bands there are gaps. In these gaps we have $\frac{1}{2}|\chi|>1$ and
so solutions of type II.

{\bf 7 Prop: The Floquet spectrum of the periodically kicked oscillator.}
The periodically kicked oscillator has a Floquet band spectrum. Each interval 
$2m\pi  \leq \omega T \leq 2(m+1)\pi,\: m=0,1,2,..$ carries two bands. The centers $T_0$ from eq. \ref{l3} are repeated  periodically, the edges given by eq.  \ref{l4}. The bands are separated  by gaps which admit only solutions of type II. 

{\em Example 8:} A simple example is provided by the choice 
$\frac{u}{2\omega}=1$. Then the half-trace $\frac{1}{2}\chi$ eq. \ref{l2} with $\beta:=\omega T$
becomes
\begin{equation}
 \label{l4a}
\frac{1}{2}\chi= \cos(\beta)+\sin(\beta) = \sqrt{2} \cos(\beta-\pi/4).
\end{equation}
The band edges $\beta^+, \beta^-$ and the band centers $\beta^0$ are located 
at
\begin{eqnarray}
 \label{l4b}
&&\beta^+: 0+2\pi m_1,\; \frac{\pi}{2} + 2\pi m_2 ,
\\ \nonumber 
&&\beta^-: \pi +2\pi m_3 ,\; \frac{3\pi}{2} + 2\pi m_4 ,
\\ \nonumber
&&\beta^0: \frac{\pi}{4} + \pi m_5 ,\: m_j= 0, \pm 1, \pm 2,...
\end{eqnarray}
Bands and gaps alternate with the same width $\frac{\pi}{2}$, as shown in 
Fig 2.
\vspace{0.4cm}

\begin{center}
\includegraphics{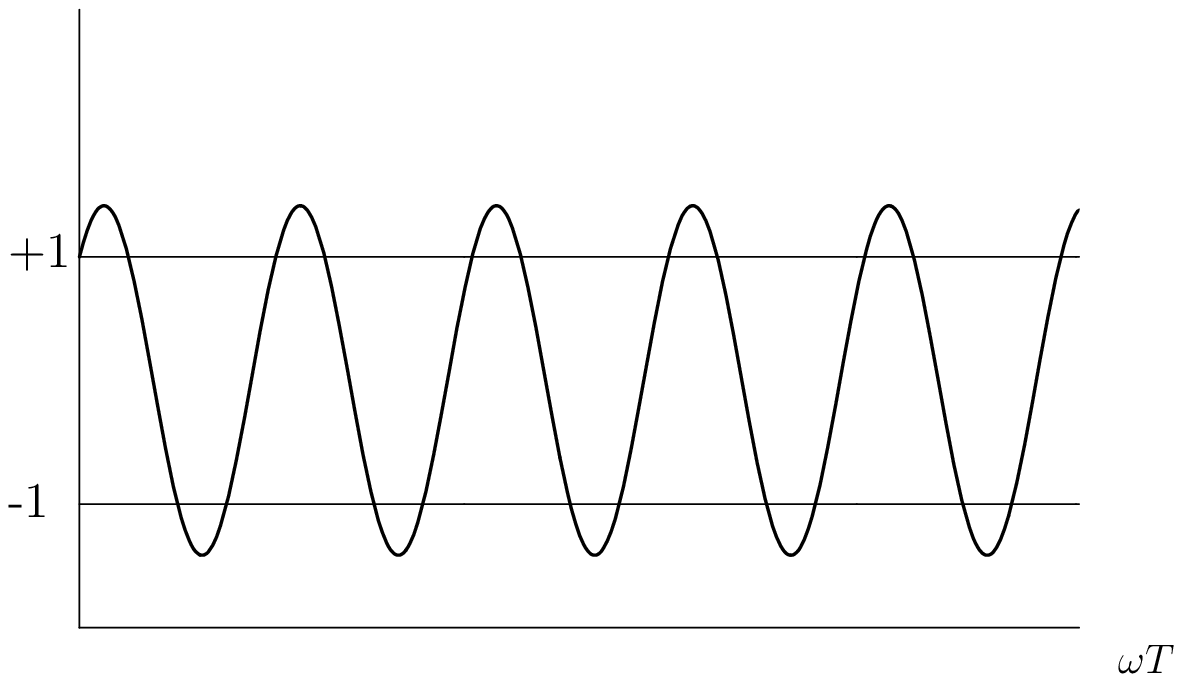} 
\end{center}

Fig. 2. {\bf Floquet band structure for the periodically kicked oscillator.} Shown is the half-trace $\frac{1}{2}\chi$ eq. \ref{l2} as a function of $\omega T$ for the parameter $\frac{u}{2\omega}=1$   of eq. \ref{l4a}.
Bands and gaps alternate periodically and  both have the width $\Delta \beta=\frac{\pi}{2}$.

\section{The quasiperiodically kicked oscillator.}

For properties of systems with a general quasiperiodic Hamiltonian 
we refer to \cite{DI}. 
The Nielsen theorem is valid for all discrete 
quasiperiodic systems of Fibonacci type, but its significance can already be seen 
on a simple example. Here we study a specific discrete quasiperiodic system of Fibonacci type in time, in analogy to a system in space from \cite{KK}. 
Following \cite{FU}, it would be worthwile to study the spectral function for
this discrete quasiperiodic system. 

For two intervals $(1,2)$ which form a quasiperiodic Fibonacci sequence we take the matrix propagators 
in the form 
\begin{equation}
\label{k1}
\lambda_i= 
\left[ 
\begin{array}{ll}
\cos(\omega T_i)& \omega^{-1}\sin(\omega T_i)\\
-\omega\sin(\omega T_i) & \cos(\omega T_i) 
\end{array}
\right]\:
\left[ 
\begin{array}{ll}
1& 0\\
u& 1 
\end{array}
\right],\;  i=1,2.
\end{equation} 
Then we compute, expanded in powers of $\frac{u}{\omega}$,  the  products 
\begin{eqnarray}
\label{k2}
&& \lambda_1\lambda_2 =
\left[
\begin{array}{ll}
 \cos(\omega(T_1+T_2))&\omega^{-1}\sin(\omega(T_1+T_2)\\
-\omega\sin(\omega(T_1+T_2))&\cos(\omega(T_1+T_2))\\
\end{array} 
\right]
\\ \nonumber 
&&+\frac{u}{\omega} 
\left[ 
\begin{array}{ll}
2\sin(\omega T_1)\cos(\omega T_2)+\cos(\omega T_1)\sin(\omega T_2)
& \omega^{-1}\sin(\omega T_1)\sin(\omega T_2)\\
\omega(2\cos(\omega T_1)\cos(\omega T_2)-\sin(\omega T_1)\sin(\omega T_2))
& \cos(\omega T_1)\sin(\omega T_2)
\end{array}
\right]
\\ \nonumber
&&+(\frac{u}{\omega})^2
\left[ 
\begin{array}{ll}
\sin(\omega T_1)\sin(\omega T_2)& 0\\
\omega\cos(\omega T_1)\sin(\omega T_2)& 0 
\end{array}
\right]
\\ \nonumber
&& (\lambda_2\lambda_1)^{-1}= 
\left[
\begin{array}{ll}
 \cos(\omega(T_1+T_2))&-\omega^{-1}\sin(\omega(T_1+T_2)\\
\omega\sin(\omega(T_1+T_2))&\cos(\omega(T_1+T_2))\\
\end{array} 
\right]
\\ \nonumber
&&+\frac{u}{\omega} 
\left[ 
\begin{array}{ll}
\sin(\omega T_1)\cos(\omega T_2)
& -\omega^{-1}\sin(\omega T_1)\sin(\omega T_2)\\
-\omega(2\cos(\omega T_1)\cos(\omega T_2)-\sin(\omega T_1)\sin(\omega T_2))
& 2\cos(\omega T_1)\sin(\omega T_2)+\sin(\omega T_1)\cos(\omega T_2)
\end{array}
\right]
\\ \nonumber 
&&+(\frac{u}{\omega})^2
\left[ 
\begin{array}{ll}
0&0\\
-\omega \sin(\omega T_1)\cos(\omega T_2)& \sin(\omega T_1)\sin(\omega T_2)\\ 
\end{array}
\right]
\end{eqnarray}
For the commutator ${\cal K}$ and its inverse we find  
\begin{eqnarray}
\label{k3}
&&{\cal K}= (\lambda_1\lambda_2)(\lambda_2\lambda_1)^{-1}
\\ \nonumber
&&= 
\left[
\begin{array}{ll}
1 & 0\\
0 & 1 \\
\end{array}
\right] 
+\frac{u}{\omega} \sin(\omega (T_1-T_2))
 \left[
\begin{array}{ll}
\cos(\omega (T_1+T_2)) & -\omega^{-1}\sin(\omega (T_1+T_2))\\
 -\omega\sin(\omega (T_1+T_2))&  -\cos(\omega (T_1+T_2))\\
\end{array}
\right] 
\\ \nonumber
&&+ (\frac{u}{\omega})^2 \sin(\omega(T_1-T_2))
\left[
\begin{array}{ll}
\sin(\omega T_1)\cos(\omega T_2) & -\omega^{-1}\sin(\omega T_1)\sin(\omega T_2) \\
\omega\cos(\omega T_1)\cos(\omega T_2)  & -\cos(\omega T_1)\sin(\omega T_2)  \\
\end{array}
\right],
\\ \nonumber 
&&{\cal K}^{-1}= (\lambda_2\lambda_1)(\lambda_1\lambda_2)^{-1}
\\ \nonumber
&&= 
\left[
\begin{array}{ll}
1 & 0\\
0 & 1 \\
\end{array}
\right] 
+\frac{u}{\omega} \sin(\omega (T_1-T_2))
 \left[
\begin{array}{ll}
-\cos(\omega (T_1+T_2)) & \omega^{-1}\sin(\omega (T_1+T_2))\\
 \omega\sin(\omega (T_1+T_2))&  \cos(\omega (T_1+T_2))\\
\end{array}
\right] 
\\ \nonumber
&&+ (\frac{u}{\omega})^2 \sin(\omega(T_1-T_2))
\left[
\begin{array}{ll}
-\cos(\omega T_1)\sin(\omega T_2) & \omega^{-1}\sin(\omega T_1)\sin(\omega T_2) \\
-\omega\cos(\omega T_1)\cos(\omega T_2)  & \sin(\omega T_1)\cos(\omega T_2)  \\
\end{array}
\right],
\end{eqnarray}
Remember that in each iteration of the Fibonacci automorphism the commutator
${\cal K}$ by eq. \ref{ff3} is transformed into its inverse. This does not affect the trace 
which becomes an invariant under the automorphism.

The half-trace of the commutator therefore becomes the invariant 
\begin{equation}
\label{k4}
{\cal I} = \frac{1}{2}{\rm Tr}({\cal K})=\frac{1}{2}{\rm Tr}({\cal K}^{-1})
= 1+ \frac{1}{2}(\frac{u}{\omega})^2 ( \sin(\omega(T_1-T_2))^2.
\end{equation}
Of interest are the points where the invariant takes the value $I=1$.
These are given from eq. \ref{k4} by 
\begin{eqnarray} 
\label{k5}
&&\sin(\omega(T_1-T_2))=0,\; 
\\ \nonumber
&&\omega(T_1-T_2)= m \pi, m= 0, \pm 1, \pm 2, \ldots.
\end{eqnarray}
For the Fibonacci case 
$T_1=T, T_2= \tau T$  we find from eqs. \ref{k4}, \ref{k5} 
\begin{equation}
\label{k6}
\omega (\tau-1)T= m \pi, m= 0, \pm 1, \pm 2, \ldots.
\end{equation}
The full commutator ${\cal K}$ at these points from eq. \ref{k3}
becomes
\begin{equation}
\label{k7}
{\cal K}= I.
\end{equation}
and we find
\newpage 
{\bf 8 Theorem}:
The invariant ${\cal I}$ as a function of $T_1, T_2$ has the value ${\cal I}=1$ at the points eqs. \ref{k5}
and \ref{k6}. These points are repeated periodically.
At each of these points, the commutator ${\cal K}$ becomes the unit matrix 
and so the two matrices eq. \ref{k1} commute.
\vspace{0.2cm}

For the Fibonacci model of eq. \ref{k6} we get periodic points where the two
matrices $\lambda_1, \lambda_2$ commute. This does not automatically imply that we have 
a quasiperiodic Floquet propagation. To assure that, we must for both 
matrices be in an individual Floquet band. The relevant parameters for the individual Floquet bands are
$\omega T, \tau \omega T$ respectively which scale by $\tau$. To find overall 
propagation we must superimpose the two band structures and 
find the regions of band overlap. 
Moreover for commutativity we must fulfill the 
condition eq. \ref{k6}. The superposition of the bands and the location 
of commutativity are shown in Fig 3. 
\vspace{0.4cm}

\begin{center}
\includegraphics{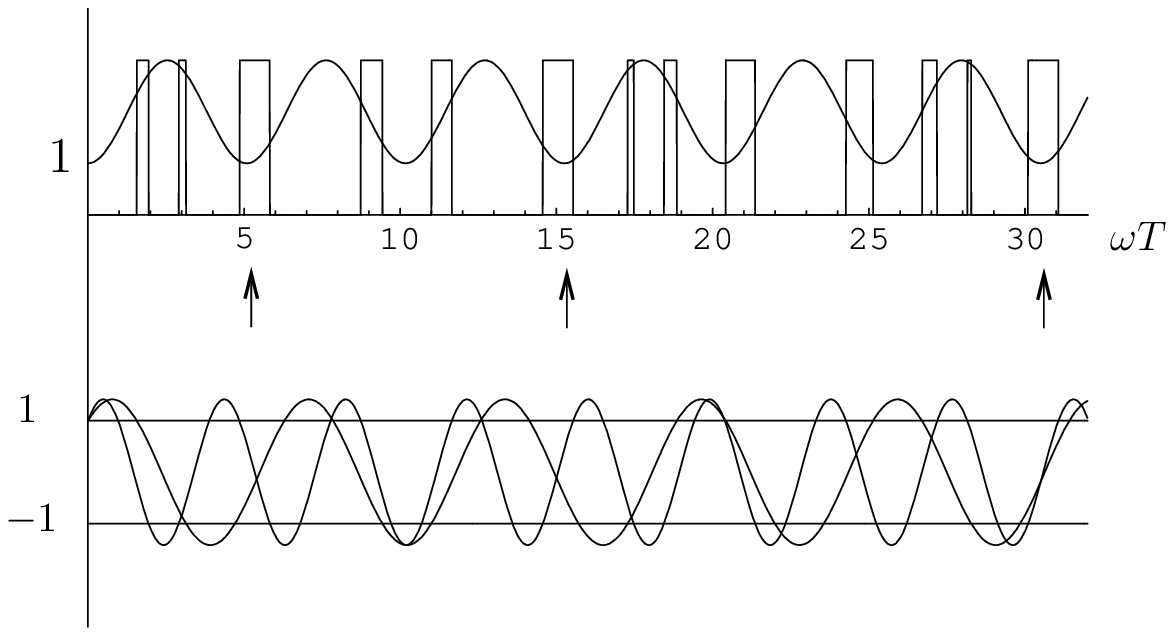} 
\end{center}

Fig. 3. {\bf Quasiperiodic Fibonacci system.} Bottom: For the parameters of eq. \ref{l4b},
the half-traces of the two systems exhibit two systems of  bands as functions of $\omega T$. Top: Overlap regions of the two systems of bands and the invariant 
${\cal I}(\omega T)$. If a commutative point with ${\cal I}=1$ hits a band overlap region, the system shows quasiperiodic Floquet behaviour. These points are marked by vertical arrows.
\vspace{0.2cm}

The propagation in phase space for multiple values of a commutative in-band case 
is shown in Fig. 4. All the points reached fall on a single ellipse equivalent to a harmonic oscillator.

\begin{center}
\includegraphics{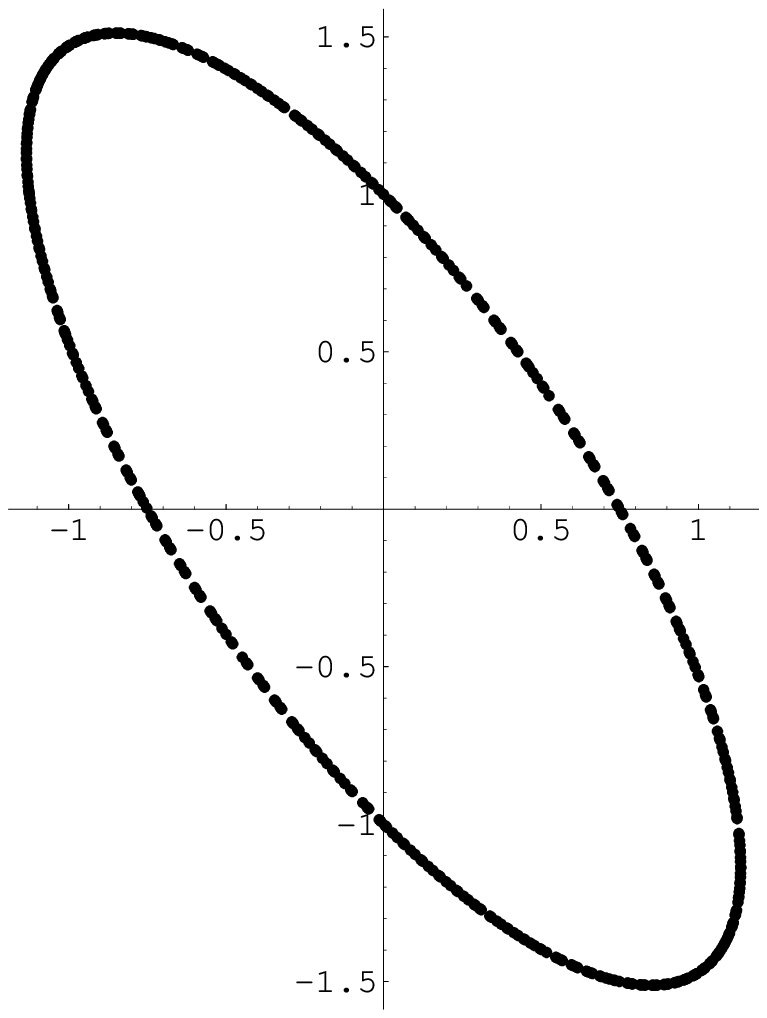} 
\end{center}

Fig. 4. Quasiperiodic Floquet theorem: For multiples of a value $\omega T$ with ${\cal I}(\omega T)=1$,
hence commutative, and moreover in the overlap of the two  Floquet bands, all phase space propagators fall on an ellipse equivalent to a harmonic oscillator.

\section{Conclusion.}
The algebraic Nielsen theorem  applied to classical and quantum quasiperiodic systems
yields a novel conserved quantity. 
This quantity as a continuous function of system parameters controls transitions between commutative and non-commutative propagation in time. The propagation of the quasiperiodically 
kicked oscillator demonstrates  an interplay between the Nielsen and 
Floquet properties.

\section*{Acknowledgement.}
One of the authors (P.K.) thanks Ch. Fulton, Dept. of  Mathematical Sciences, Florida Institute of Technology, Melbourne, Florida, for helpful comments. Fulton  pointed out reference \cite{DI} for quasiperiodic Hamiltonians, and the need to study the spectral function for quasiperiodic Sturm-Liouville problems.

\section*{Appendix A: Irreps of the discrete translation group.}

We give here the orthogonality and completeness relations for the discrete
translation group $C_{\infty}$ as discussed in \cite{PK3}. In the present case they 
apply to Floquet states.

The orthogonality relation is
\begin{eqnarray}
\label{A1}
&&D^{\Omega}(nT)= \exp( 2\pi i\Omega nT),
\\ \nonumber
&&\Omega \in BZ=\{ 0, 2\pi/T\},\; n \in \{0, \pm1, \pm 2,...\}
\\ \nonumber
&&\sum_{n} 
\overline{D^{\Omega}}(nT) D^{\Omega'}(nT) d\Omega= 
|BZ| \delta(\Omega-\Omega').
\end{eqnarray}
Eq. \ref{A1} yields an orthogonality relation between evolutions belonging to
two different Floquet indices. It does not imply an orthogonality between
pairs with the same Floquet index from different Floquet bands,
as they are found for example in section 10.  
The completeness relation is
\begin{equation}
\label{A2}
\int_{\Omega \in BZ} 
\overline{D^{\Omega}}(nT) D^{\Omega}(mT) d\Omega= 
|BZ| \delta_{n,m}.
\end{equation}

\section*{Appendix B: Classical propagator and transfer matrix.}

The matrix which propagates the canonical pair of position and momentum 
$\{X(t), P(t)\}$ we denote as the classical (symplectic) propagator  
$g(t) \in Sp(2,R),\; det(g(t))=1$. 
The matrix which propagates a fundamental pair   of 
position  and velocity $\{X(t), \dot{X}(t)\}$ we denote as the transfer
matrix $M(t)$. Again we may choose $det(M(t))=1$. For a hamiltonian 
with standard kinetic energy we have the relation $P(t)=m \dot{X}(t)$. 
>From this we get the conjugation relation
\begin{equation}
\label{B1}
M(t) = 
\left[ 
\begin{array}{ll}
\sqrt{m}&0\\
0&\frac{1}{\sqrt{m}}
\end{array}
\right]
g(t) 
\left[ 
\begin{array}{ll}
\frac{1}{\sqrt{m}}&0\\
0&\sqrt{m} 
\end{array}
\right]
\end{equation}
between the symplectic propagator $g(t)$ and the transfer matrix
$M(t)$. From the relation eq. \ref{B1} it follows that for a sequence of strings
both their symplectic propagators and their transfer matrices are
multiplied in the same order.

The first and second  column of the propagator matrix $g(t)$ each 
yield a fundamental  canonical pair $\{I, II\}$ of 
solutions and trajectories for the
hamiltonian equations of motion. For quadratic hamiltonians, 
the hamiltonian
equations of motion are linear. Then the most general classical solution
is a linear superposition of  fundamental solutions.
The superposition changes the classical energy which therefore must
be recalculated for any superposition.


\begin{thebibliography}{99}

\bibitem{AR} Arnold V I, 
{\em Mathematical Methods of classical mechanics},
Springer, Berlin (1978)

\bibitem{BA} Bargmann V, {\em Group representations in Hilbert spaces 
of analytic functions} (1968)

\bibitem{BY} Bayfield, J E,
{\em Quantum evolution},
Wiley, New York (1999)

\bibitem{DO}
Dodonov V V and Man'ko V I, {Invariants and evolution of non-stationary 
quantum systems}, ed. Markov, Nova Science Publishers, Cormack, N.Y. (1989)

\bibitem{DI} Dinaburg E and Sinai Ya,
{\em The One-dimensional Schr\"odinger Equation with a Quasiperiodic Potential},
Functional Anal. Appl, {\bf 9} (1975) 279-290 

\bibitem{FU} Fulton Ch, Pearson D, and Pruess, St,
{\em Computing the spectral function for singular Sturm-Liouville problems},
J Comp. Appl. Math. {\bf 176} (2005) 131-162


\bibitem{ML} Lothaire M, {\em Combinatorics on words}, 
Addison-Wesley, Reading (1983)

\bibitem{KA1} Karner G, Man'ko V I, and Streit L, {\rm Quasi-energies, loss-energies and stochasticity}, Reports on Mathematical Physics, {\bf 29} (1991) 177-93

\bibitem{KA2} Karner G, Man'ko V I, and Streit L, Proc. Lebedev Physical Institute Nauka {\bf 208}
(1992) 226

\bibitem{TK1} Kramer T, {\em Quantum ballistic motion in uniform 
electric and magnetic fields}, Diploma thesis, TU Munich (2000)

\bibitem{TK2} Kramer T, {\em Matter waves from localized sources 
in homogeneous force fields}, PhD thesis, TU Munich, 2003.
Online: http://tumb1.biblio.tu-muenchen.de/publ/diss/ph/2003/kramer.html

\bibitem{KK} Kramer P and Kramer T,
{\em Exact electron states in 1D (quasi-) periodic arrays of 
$\delta$-potentials}, in: {\em From quasicrystals to more complex systems}
eds. F Axel et al, Springer, Berlin 2000, pp. 85-114.

\bibitem{PK1} Kramer P, 
{\em Energy gauge and electron confinement in quasicrystals.}
J Phys {\bf A 31} (1998) 743-56

\bibitem{PK2} Kramer P,
{\em Algebraic structure for one-dimensional quasiperiodic systems}
J Phys {\bf A 26} (1993) 213-28

\bibitem{PK3} Kramer P and Lorente M,
{\em Discrete and continuous symmetry via induction and duality},
in: Symmetries in Science {\bf X}, B Gruber and M Ramek eds.,
Plenum, New York 1998, pp. 165-77

\bibitem{PK4} Kramer P,
{\em Quasiperiodic systems},
in: Encyclopedia of Mathematical Physics,
eds. J P Francoise, G L Naber, Sh Tsun Tsou,
Elsevier, Amsterdam  2006, 308-315

\bibitem{KO}
Kohmoto M, Kadanoff L P and Tang C,
Phys. Rev. Lett. {\bf 50} (1983) 1870





\bibitem{MA} W Magnus, A Karrass, D Solitar,
{\em Combinatorial group theory}, Dover, New York 1976

\bibitem{MM} Malkin I A and Man'ko V I,
{\em Dynamical symmetries and coherent states of quantum systems},
Nauka, Moscow  (1979) (in Russian) 


\bibitem{MQ} Moshinsky M and Quesne C, 
{\em Linear canonical transformations and their unitary 
representation}, J Math Phys {\bf 12} (1971) 1772-





\bibitem{WM} Winternitz P and Moshinsky M,
{\em Quadratic hamiltonians in phase space and their eigenstates},
J Math Phys {\bf 21} (1980) 1667-82  

\end{thebibliography}
\end{document}